\documentclass[aps,prb,reprint,superscriptaddress]{revtex4-2}

\usepackage{graphicx}
\usepackage{dcolumn}
\usepackage{bm}
\usepackage{xcolor}

\begin{document}

\title{What causes the variation in superconducting properties of UTe$_{2}$?}

\author{C. S. Kengle}
\affiliation{Los Alamos National Laboratory, Los Alamos, NM 87545}

\author{Noah Schnitzer}
\affiliation{Department of Materials Science and Engineering, Cornell University, Ithaca, NY, 14853, USA}
\affiliation{Kavli Institute at Cornell for Nanoscale Science, Ithaca, NY, 14853, USA}

\author{M. M. Bordelon}
\affiliation{Los Alamos National Laboratory, Los Alamos, NM 87545}

\author{S. M. Thomas}
\affiliation{Los Alamos National Laboratory, Los Alamos, NM 87545}

\author{P. F. S. Rosa}
\affiliation{Los Alamos National Laboratory, Los Alamos, NM 87545}

\date{\today}

\begin{abstract}
Reaching a consensus on the superconducting order parameter of unconventional superconductors remains a central challenge in the field of magnetically-mediated superconductivity. Though UTe$_{2}$ is largely accepted as a rare example of an odd-parity superconductor, its precise order parameter remains highly debated, even at ambient conditions. A key underlying issue is the large sample-to-sample variation in superconducting properties at zero applied pressure and magnetic field. Here, we investigate the origin of the observed variation by means of single crystal x-ray diffraction  (SC-XRD) and scanning transmission electron microscopy (STEM) measurements. Our results reveal highly ordered crystalline lattices, in agreement with the expected $Immm$ structure, and no signs of uranium vacancies. Tiny amounts of interstitial defects, however, are observed on the Te2 layers that host Te chains along the $b$ axis. We argue that these defects give rise to slightly enhanced atomic displacement parameters observed in SC-XRD data and are enough to disrupt the unconventional superconducting state in UTe$_{2}$. Our findings highlight the need to focus future  order parameter determination efforts on single crystals of UTe$_{2}$ with minimal amounts of structural disorder.
\end{abstract}

\maketitle

\section{Introduction}

In phonon-mediated superconductors, the fully-gapped superconducting (SC) order parameter ($\psi$) is well described by the Bardeen-Cooper-Schrieffer (BCS) theory wherein $\psi$ displays even $s$-wave symmetry with a spin-singlet configuration and vanishing orbital angular momentum \cite{Bardeen1957,tinkham2004}.  In magnetically-mediated unconventional superconductors, however, strong on-site repulsion drives electrons to pair in less symmetric channels with higher angular momentum (e.g., odd-parity $p$-wave or even-parity $d$-wave), and nodal SC gaps are typically favored \cite{scalapino2012,stewart2017}. The precise determination of the SC order parameter in unconventional superconductors is not only crucial for understanding materials behavior as a function of external tuning parameters (e.g., magnetic fields, disorder), but it is also a prerequisite for applications based on exotic quasiparticle excitations \cite{sato2017}.

Depending on the crystal lattice, many different order parameters may be allowed by symmetry \cite{Sigrist1991}, which poses a challenge to experimental investigations \cite{Maeno2024}. For example, the orthorhombic lattice of unconventional superconductor UTe$_{2}$ (space group 71, $Immm$) gives rise to four possible even-parity irreducible representations, $A_{g}$ and $B_{ig}$ $(i=1,2,3)$, and four odd-parity representations, $A_{u}$ and $B_{iu}$ $(i=1,2,3)$~\cite{aoki2022}. In addition, different representations can be paired into various multi-component states. Two features strongly point to an odd-parity SC state in UTe$_{2}$: the remarkably large upper critical field in UTe$_{2}$~\cite{lewin2023review}, well above the Pauli limit expected for an even-parity configuration, and the small Knight shift drop at the SC transition temperature, $T_{c}$, in nuclear magnetic resonance measurements \cite{Matsumura2023}. Though this result constrains the allowed representations $A_{u}$, $B_{1u}$, $B_{2u}$, and $B_{3u}$, dozens of different single and multi-component proposals have been put forward to explain the SC state of UTe$_{2}$ at ambient conditions~\cite{metz2019point,ishizuka2021periodic,hayes2021multicomponent,machida2021nonunitary,shaffer2022chiral,yu2023theory,lee2023anisotropic,kittaka2020orientation,bae2021anomalous,ishihara2023chiral,iguchi2023microscopic,tei2024pairing,hazra2024pair,roising2024thermodynamic,christiansen2025quasiparticle,yamazaki2025higher,theuss2024single}. 

A key long-standing issue is whether UTe$_{2}$ intrinsically displays one or two SC transitions at ambient conditions. Because all irreducible representations are one dimensional in the orthorhombic $D_{2h}$ space group, the initial observation of a double $T_{c}$ in specific heat measurements implied a multicomponent state~\cite{hayes2021multicomponent}. However, most recent experiments on high-quality single crystals, grown through either the chemical vapor transport (CVT) or the molten-salt flux (MSF) techniques, reveal a residual resistivity ratio (RRR) higher than 40 and single-component superconductivity in the bulk of UTe$_{2}$ at ambient conditions~\cite{aoki2022}. The experimental probes that argue for a single transition include specific heat, uniaxial strain, ultrasound, polar Kerr, scanning SQUID, muon spin resonance, nuclear magnetic resonance, thermal conductivity, and scanning tunneling microscopy 
measurements~\cite{cairns2020,rosa2022,weiland2022,sakai2022,aoki2024molten,girod2022thermodynamic,theuss2024single,ajeesh2023fate,Matsumura2023,suetsugu2024fully,hayes2025robust,gu2025pair,sharma2025observation}. These different techniques, however, do not agree on the exact irreducible representation at play, and reports exist that argue for either a fully-gapped $A_{u}$ state or different nodal $B_{iu}$ configurations. In addition, a few experimental techniques argue for a multi-component superconducting state, including penetration depth ($B_{3u}+iA_{u}$), angle-dependent specific heat, and microwave surface impedance measurements~\cite{kittaka2020orientation,ishihara2023chiral,bae2021anomalous}.

An outstanding question is therefore the origin of the variation in SC properties in UTe$_{2}$ and whether underlying structural disorder could be causing the discrepancy between different experimental results.
Here we present an investigation of representative superconducting UTe$_{2}$ single crystals through single crystal x-ray diffraction (SC-XRD) and scanning transmission electron microscopy (STEM) measurements. Overall, our results are consistent with a highly ordered lattice crystallizing in the expected orthorhombic $Immm$ space group, and we observe no signs of uranium vacancies, which are known to exist in nonsuperconducting samples. Tiny amounts of interstitial defects, however, become evident in both STEM maps and as additional charge density in SC-XRD refinements. These interstitials are observed primarily on the Te2 layers, which host Te chains along the $b$ axis. We argue that these defects give rise to slightly enhanced atomic displacement parameters observed in SC-XRD data and are enough to disrupt the unconventional SC state in UTe$_{2}$ because Te2 $p$ bands are the primary contribution to the band structure at the Fermi level~\cite{miao2020,christovam2024}. Our findings highlight the need to focus future order parameter determination efforts on crystals of UTe$_{2}$ with minimal amounts of structural disorder.

\section{Results}

\begin{figure}[!h]
	\includegraphics[width=\columnwidth]{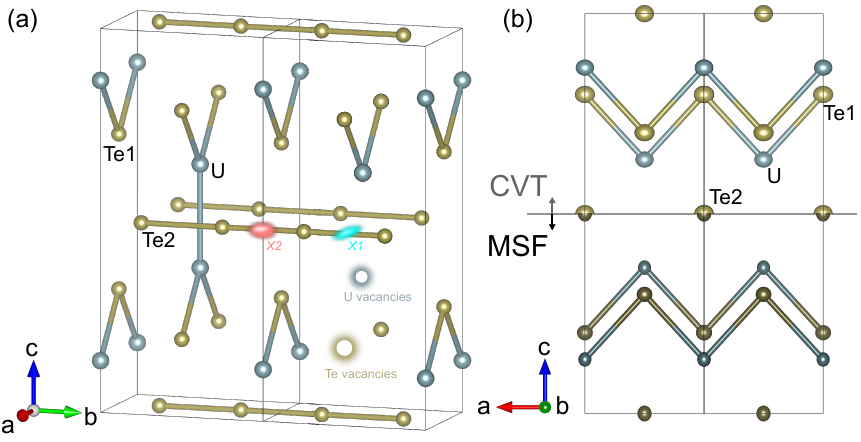}
	\caption{(a) Orthorhombic crystal structure of UTe$_{2}$ and its possible crystallographic defects. (b) Illustration of atomic displacement parameter for each atom, which is represented by its 99\% probability ellipsoid.}
	\label{fig:structure}
\end{figure}

We start by discussing potential crystallographic defects in the orthorhombic crystal structure of UTe$_{2}$. The most obvious defect is the presence of uranium vacancies (see Fig.~\ref{fig:structure}a), and nonsuperconducting crystals (RRR $\lesssim 2$) clearly hold vacancy concentrations up to $5$\% according to two independent SC-XRD analyses \cite{rosa2022, haga2022}. One report also argued for the presence of Te vacancies based on energy dispersive x-ray measurements~\cite{cairns2020}; however, SC-XRD measurements have not confirmed this assessment. As detailed below, the present work reveals an important additional structural defect in SC samples: interstitials around Te2 atoms (see $X_{1}$ and $X_{2}$ in Fig.~\ref{fig:structure}a). 

Importantly, the effects of crystallographic disorder can be also detected in the atomic displacement parameters (ADPs)~\cite{rosa2022,weiland2022}, which reflect the mean-square displacement of atoms about their equilibrium positions and encompass information about both vibrations and static disorder. The ADPs for both CVT- and MSF-grown crystals at room temperature are illustrated in Fig.~\ref{fig:structure}b and reveal larger magnitudes in CVT-grown samples, which indicate enhanced static disorder.

\begin{figure}[!h]
	\includegraphics[width=0.8\columnwidth]{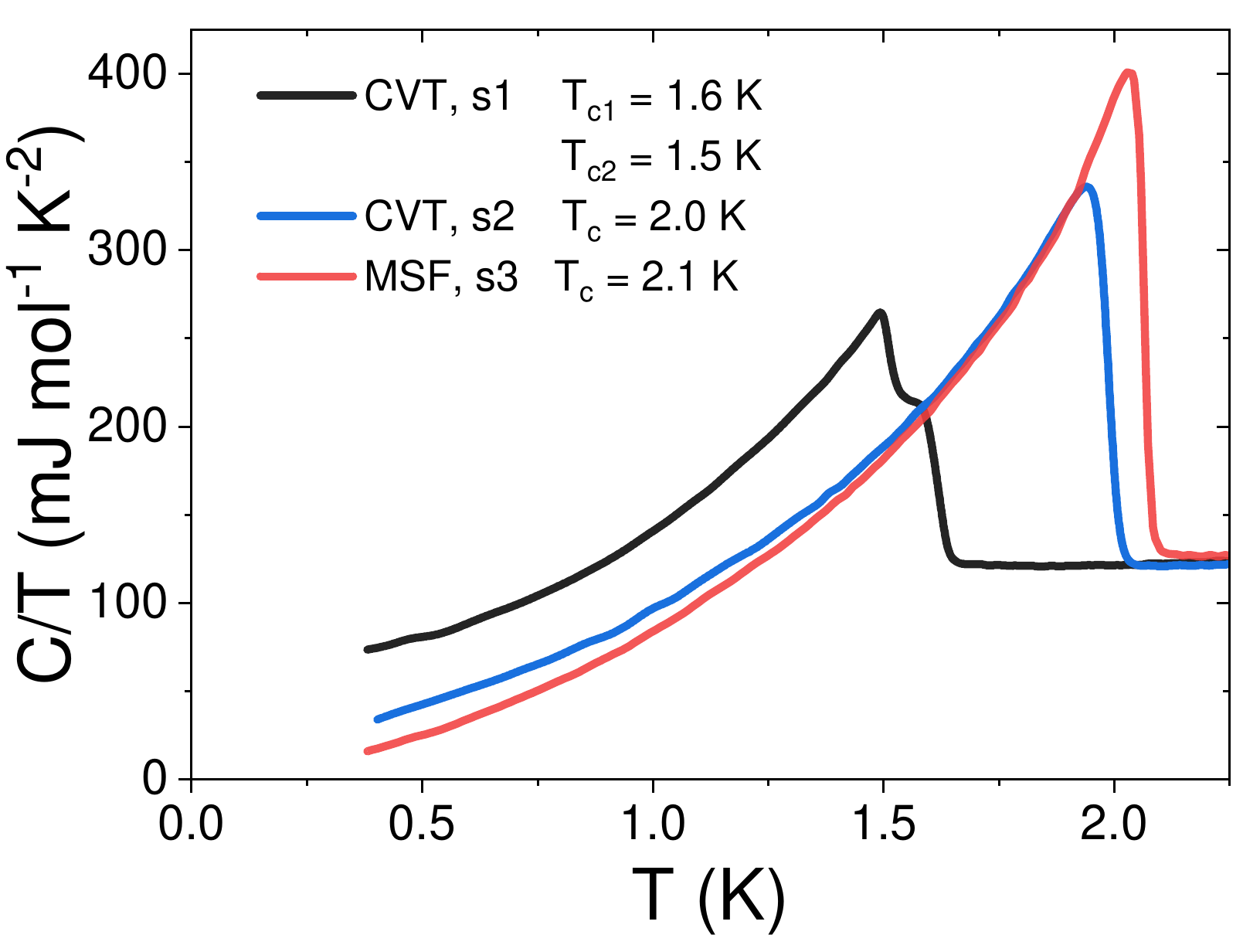}
	\caption{Specific heat, $C/T$, as a function of temperature for three representative UTe$_{2}$ crystals investigated in this work.}
	\label{fig:Cp}
\end{figure}

In the following, we will discuss three representative UTe$_{2}$ single crystals: a CVT-grown sample with two SC transitions at $T_{c1}=1.6$~K and $T_{c2}=1.5$~K (s1, RRR $=30-40$); a CVT-grown sample with a single SC transition at $T_{c}=2.0$~K (s2, RRR $= 90$); and an MSF-grown sample with a single SC transition at $T_{c}=2.1$~K (s3, RRR $= 270$). Figure~\ref{fig:Cp} shows the specific heat, $C/T$, as a function of temperature for all three samples. The MSF-grown sample displays the lowest residual specific heat coefficient, $\gamma_{\mathrm{s3}} = 5.5~\mathrm{mJ/mol.K}^{2}$, compared to CVT-grown samples, $\gamma_{\mathrm{s2}} = 24.2~\mathrm{mJ/mol.K}^{2}$ and $\gamma_{\mathrm{s1}} = 61.1~\mathrm{mJ/mol.K}^{2}$, in agreement with previous reports~\cite{aoki2022}. For the most recent discussion on UTe$_{2}$ crystal growth techniques, we refer the reader to \cite{aoki2024molten}.

\begin{figure*}[!ht]
	\includegraphics[width=1\textwidth]{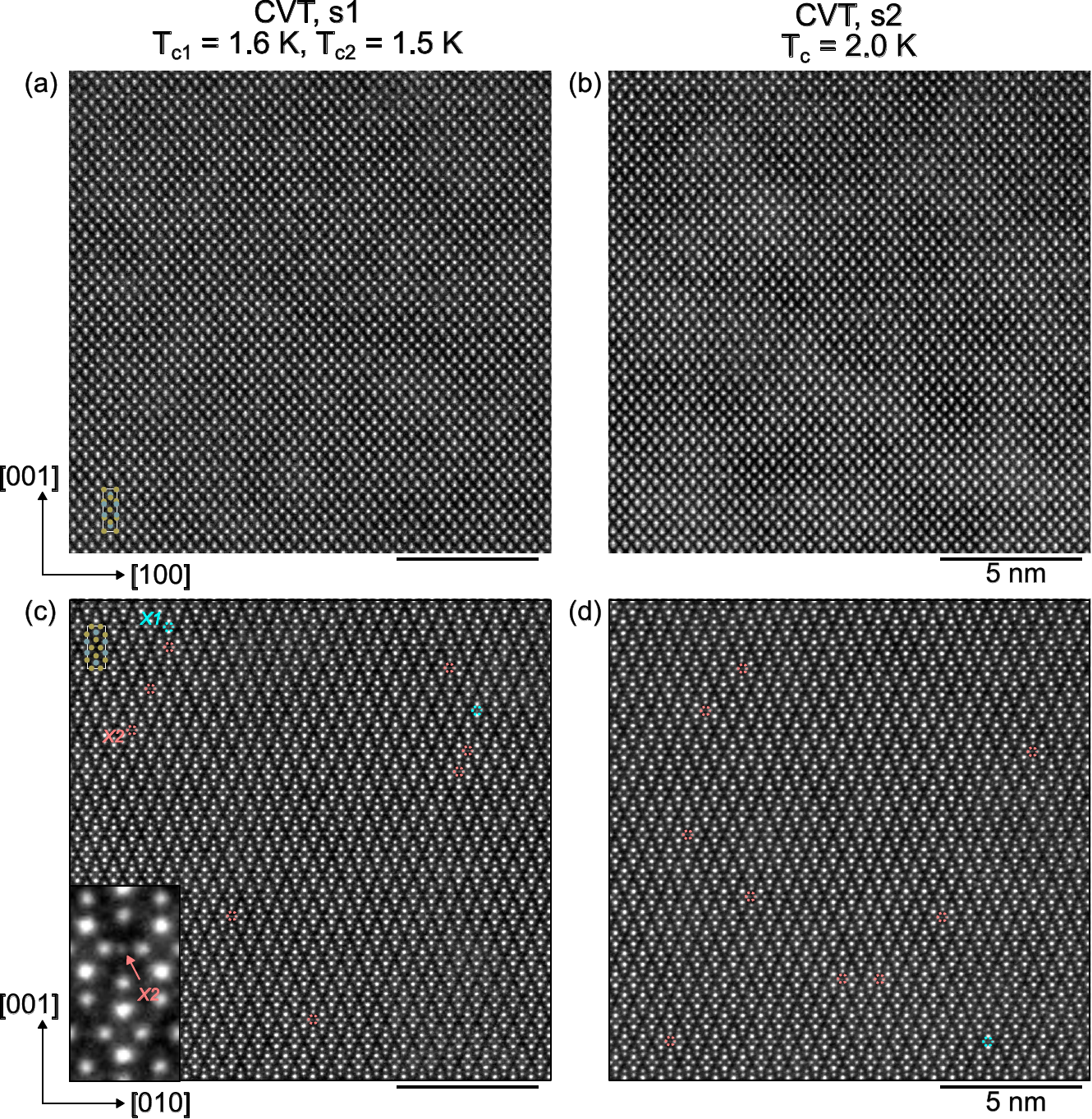}
	\caption{(a) High-angle annular dark-field scanning transmission electron microscopy  (HAADF-STEM) image on the [010] zone axis of sample s1. Inset in the bottom left shows the unit cell confirmed by single crystal x-ray diffraction measurements. (b) HAADF-STEM image on the [010] zone axis of sample s2. (c) HAADF-STEM image on the [100] zone axis of sample s1. Circles mark two types of interstitials on the Te2 layer: X1 (blue) and X2 (pink). Inset in the bottom left expands a unit cell with the position of the X2 interstitial marked by an arrow. (d) HAADF-STEM image on the [100] zone axis of sample s2, which shows a similar distribution of occupied interstitial sites.}
	\label{fig:TEM-010}
\end{figure*}

We now turn to high-angle annular dark-field STEM (HAADF-STEM) imaging of samples s1 and s2. Figures~\ref{fig:TEM-010}a-b show HAADF-STEM images of s1 and s2, respectively, on the [010] zone axis, which directly visualize the atomic positions in the $ac$ plane of UTe$_{2}$. The crystalline structures of both samples are very similar and match closely the expected $Immm$ structure. On this axis, uranium vacancies, line defects, and grain boundaries are not observed in our STEM images, which confirms the high degree of crystallinity of both samples.

HAADF-STEM images on the [100] zone axis, however, reveal subtle deviations from the ideal $Immm$ structure. Figures~\ref{fig:TEM-010}c-d show HAADF-STEM images of s1 and s2, respectively, on the [100] zone axis, which directly visualize the atomic positions in the $bc$ plane of UTe$_{2}$. In both samples, two interstitials are found primarily on the Te2 layer: $X_{1}$ (blue circles) and $X_{2}$ (pink circles). The inset of Figure~\ref{fig:TEM-010} shows a zoomed-in view of $X_{2}$, whereas Fig.~\ref{fig:structure} shows a schematic illustration of the different locations. Because HAADF-STEM images on the $ac$ plane did not detect an obvious contribution from point defects, our resuts suggest that the interstitials are predominantly located between Te2 atoms along the $b$ axis. Our data, however, cannot rule out small shifts along either $a$ or $c$ axes.  The seemingly random spatial distribution of defects suggest that these interstitials are just point defects, rather than evidence of a planar defect along the projection axis. We note that the defect density is similar in both samples, and the detection of any systematic differences between samples is hindered by the lack of statistics: the interstitials are very sparse and their intensities very low.

\begin{figure*}[!ht]
	\includegraphics[width=0.9\textwidth]{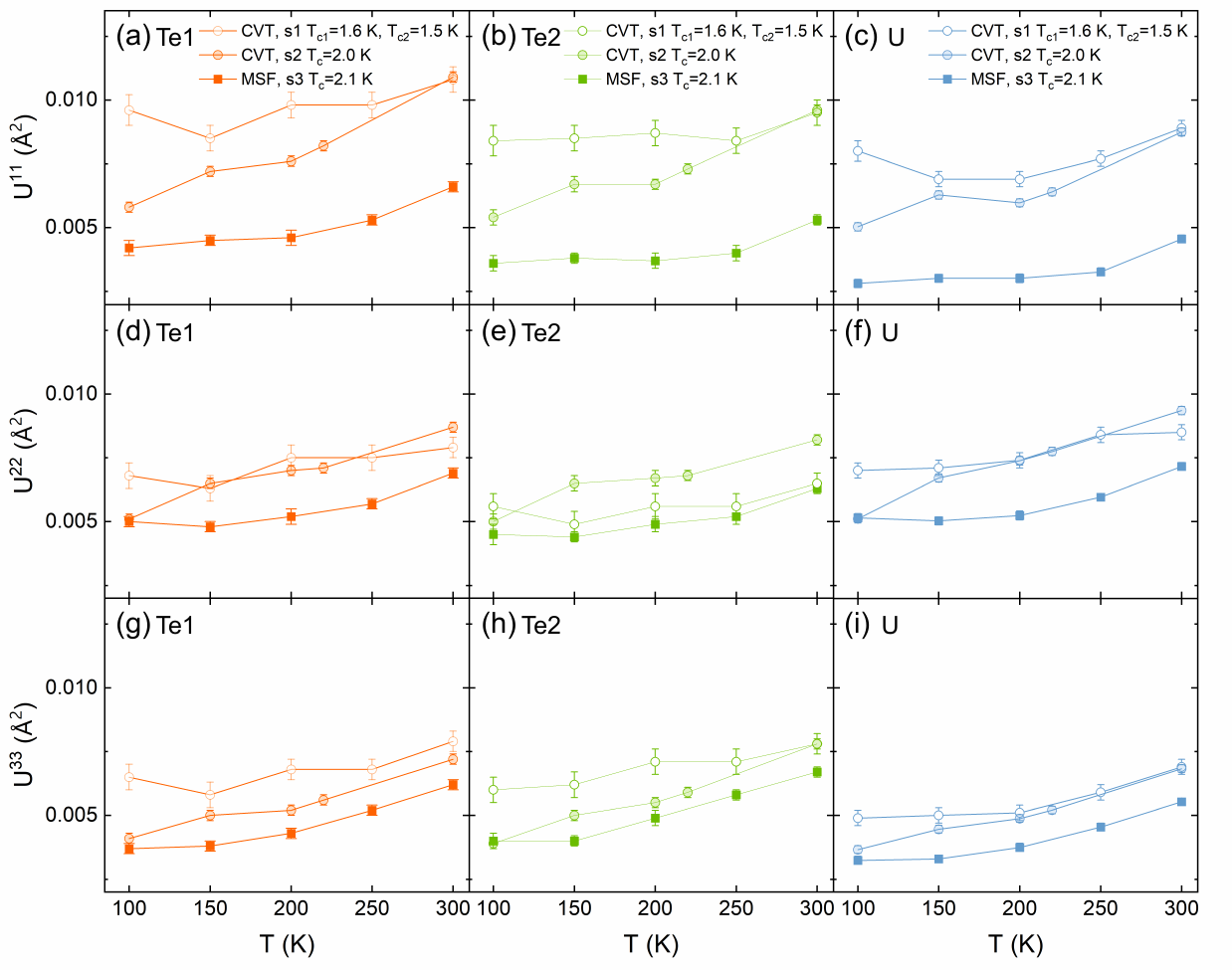}
	\caption{Anisotropic displacement parameters for samples s1 (empty circles), s2 (shaded circles), and s3 (solid squares) for each of the three unique atomic positions. (a-c) The displacement along the $a$ axis, $U_{11}$, for Te1, Te2, and U, respectively. (d-f) The displacement along the $b$ axis, $U_{22}$, for Te1, Te2, and U, respectively. (g-i) The displacement along the $c$ axis, $U_{33}$, for Te1, Te2, and U, respectively. All axes have the same scale.}
	\label{fig:ADP}
\end{figure*}

To shed further light on the role of static disorder in UTe$_{2}$, we turn to SC-XRD measurements, which provide information about the lattice parameters, atomic positions, and charge density of a material. Remarkably, the lattice parameters of samples s1, s2, and s3 are identical within experimental uncertainty (see Table~\ref{tab:parameters}). In addition, the fractional atomic coordinates are also indistinguishable, and all three sites (U, Te1, and Te2) are fully occupied. A comparison between the isotropic displacement parameters, $U_{\mathrm{iso}}$, between samples s1 and s3, however, shows hints of a trend (see Table~\ref{tab:ISO}). For example, $U_{\mathrm{iso}}$ for the Te2 atom is 30~\% larger in sample s1 than that in sample s3, which indicates an enhancement in static disorder in sample s1.

    \begin{table}

      \centering
  
      \begin{tabular}{cccccc}
  
      \hline \hline
  
        Sample & a & b & c & V \\
  
         & $\mathrm{\AA}$ &  $\mathrm{\AA}$ &  $\mathrm{\AA}$ &  $\mathrm{\AA}^3$ \\ \hline
  
         s1, CVT & 4.1612(7)  & 6.1315(19) & 13.970(5) & 356.45(18)  \\ \hline
  
         s2, CVT & 4.1625(11) & 6.1342(11) & 13.978(2) & 356.91(13) \\ \hline
  
         s3,  MSF & 4.1623(11) & 6.1347(11) & 13.979(3) & 356.96(13) \\ \hline \hline
  
      \end{tabular}
  
      \caption{Lattice parameters of UTe$_{2}$ at 300~K.}
  
      \label{tab:parameters}
  
  \end{table}

  \begin{table}

    \centering

    \begin{tabular}{cccc}

    \hline \hline

       Sample & Te1 U$_{iso}$ & Te2 U$_{iso}$ & U U$_{iso}$ \\

        &  $\mathrm{\AA}^2$&  $\mathrm{\AA}^2$ &  $\mathrm{\AA}^2$\\ \hline

       s1, CVT  & 0.0089(3)     & 0.0079(3)     & 0.0081(2) \\ \hline

       s3, MSF  & 0.00656(12) & 0.00610(13) & 0.00575(11)\\ \hline \hline

    \end{tabular}

    \caption{Isotropic displacement parameters of UTe$_{2}$ at 300~K.}

    \label{tab:ISO}

\end{table}

    Figure~\ref{fig:ADP} shows the complete set of anisotropic displacement parameters for U, Te1, and Te2 as a function of temperature. On cooling, all ADPs tend to decrease monotonically, as expected from reduced thermal vibrations.
    Notably, a trend emerges independent of temperature: the magnitude of most ADPs increases systematically from sample s3 (MSF, $T_{c}=2.1$~K) to sample s2 (CVT, $T_{c}=2.0$~K) to sample s1 (CVT, lower split $T_{c}$). More importantly, our results clearly indicate that the MSF-grown crystal has the lowest amount of static disorder, in agreement with its high RRR and low residual specific heat. Remarkably, the $a$-axis ADP, $U_{11}$, displays the clearest difference between samples: $U_{11}$ in sample s1 is about 60\% larger than $U_{11}$ in sample s3. Given our STEM findings of interstitials primarily along the Te2 chains along $b$, one would naively expect larger ADPs for the Te2 atoms and larger contributions along the $b$ axis. Our SC-XRD results, however, do not agree with this naive expectation and therefore suggest a nontrivial response of atomic motion to static disorder along $b$.

More information about the nature of the interstitials can be obtained from SC-XRD refinements. Though the refinement residuals are small ($R_{1} \sim 2.5$\%), we observe minute signs of additional charge density. In particular, the largest density peak that does not fit the expected $Immm$ structure is located at $X_{1}=(0.3, 0, 0.54)$ in sample s1 and $X_{1}=(0, 0, 0.56)$ in samples s2 and s3. Though there is a difference in the $a$-axis coordinate, both $X_{1}$ peaks are positioned around to the Te2 chains (see $X_{1}$ region in Fig.~\ref{fig:structure}a), akin to our STEM findings. We note, however, that interstitials in SC-XRD appear more strongly in the $X_{1}$ region, whereas STEM images detect the majority of interstitials around the $X_{2}$ region (see  Fig.~\ref{fig:structure}a), a difference that we attribute to a combination of statistics and the nature of the different techniques utilized here.

\section{Discussion}

Though a one-to-one comparison between SC-XRD and STEM is hindered by the inherent differences between these two techniques, our combined results point to a subtle structural disorder primarily on the Te2 layers of UTe$_{2}$. More specifically, interstitials are directly visualized by STEM around the Te2 atoms. SC-XRD concurrently observes additional charge density, also around Te2 atoms, and a systematic decrease in atomic displacement parameters with sample quality.

The next logical question is the nature of the observed interstitials. Because the distance between the interstitial atom ($X$) and the Te2 atom is very short ($\sim 2$~\AA), we first consider the possibility of light elements being incorporated into
the structure during the growth process. In particular, carbon, nitrogen, and oxygen are known impurities in depleted uranium that could give rise to such short Te-$X$ distances. HAADF-STEM imaging, however, is only sensitive to high-Z elements, and our observations thus indicate that the observed interstitials are likely not C, N, or O, but rather a higher-Z element.
Though further investigations will be required to precisely determine the nature of the interstitials in UTe$_{2}$, we hypothesize that copper (a common impurity in tellurium) is the most likely interstitial at play here.

More broadly, interstitial defects may play an important role in other phenomena observed in UTe$_2$ at ambient conditions.
For example, the charge density wave (CDW), which is observed in surface-sensitive scanning tunneling microscopy (STM) measurements, is absent in bulk measurements \cite{aishwarya2023magnetic,gu2023detection,lafleur2024inhomogeneous,kengle2024absence_a,kengle2024absence_b,theuss2024absence}. This discrepancy leads to questions about differences between surface and bulk critical behavior \cite{szabo2025intertwiningbulksurfacecase}.
Recent reports confirm the surface nature of the CDW phase and find that the largest CDW amplitudes tend to be pinned at defects~\cite{talavera2025surfacechargedensitywave}.
Surface pinning of a CDW via defects has been observed in, e.g., Ca$_{2-x}$Na$_x$CuO$_2$Cl$_2$ \cite{hanaguri2004checkerboard} and used to argue for the presence of an extraordinary phase transition \cite{brown2005surface, 1983_Binder_PhaseTransCritPhen}. 

We note that the presence of defects could also shed light into other superconducting states as a function of applied pressure or magnetic field. For example, the reentrant superconducting phase induced by fields applied along the [011] direction is remarkably robust against disorder and survives even in samples that are not superconducting at ambient conditions~\cite{Wu2024,Frank2024}. In addition, electrical resistivity measurements under pressure indicate that the pressure-induced superconducting phase above 0.3~GPa is also more robust against disorder compared to the ambient-pressure superconducting state~\cite{Ajeesh2024}.

Finally, one might wonder how tiny amounts of interstitials around Te2 atoms can have such dramatic effects on the superconducting state of UTe$_{2}$, which is driven by $5f$ physics. We argue that this is neither surprising nor unprecedented. First, unconventional superconductors are known to be highly susceptible to nonmagnetic disorder. In fact, small amounts of aluminum incorporated into the cubic structure of unconventional superconductor UBe$_{13}$ have been shown to suppress the superconducting transition temperature and even give rise to an apparent double transition in some samples~\cite{amon2018tracking}. In addition, band structure calculations and photoemission measurements in UTe$_{2}$ clearly show that the dominant non-$f$ contribution at the Fermi level ($E_{F}$) comes from highly-dispersive Te2 $p$ bands~\cite{miao2020,christovam2024}. Even small changes to these $p$ bands are therefore expected to impact the resulting hybridization with $5f$ bands at $E_{F}$.

\textit{Note added} -- During the development of this work, we became aware that Svanidze \textit{et al.} performed a similar study of UTe$_{2}$ single crystals by x-ray diffraction and transmission electron microscopy~\cite{Svanidze2025}. Our experimental findings are in general agreement; however, Svanidze \textit{et al.} argue for the presence of local deviations from the translational symmetry of the main atomic arrangement in UTe$_{2}$, namely a superposition of similar lattices with different orientations, that we do not seem to observe here.

\section{Conclusions}
We report a systematic structural investigation of UTe$_{2}$ single crystals through scanning transmission electron microscopy and x-ray diffraction. We focus on three representative samples: a CVT-grown sample with two SC transitions at $T_{c1}=1.6$~K and $T_{c2}=1.5$~K; a CVT-grown sample with a single SC transition at $T_{c}=2.0$~K, and an MSF-grown sample with a single SC transition at $T_{c}=2.1$~K. Our results reveal highly ordered crystalline lattices, in agreement with the expected $Immm$ structure, and no signs of uranium vacancies, line defects, or grain boundaries. Tiny amounts of interstitial defects, however, are observed around the Te2 layers that host Te chains along the $b$ axis. We argue that these defects give rise to slightly enhanced atomic displacement parameters observed in SC-XRD data, a strong indication of static disorder, and are enough to disrupt the unconventional superconducting state in UTe$_{2}$. Our results reveal that the molten-salt-grown crystal has the lowest amount of static disorder, in agreement with its highest residual resistivity ratio and lowest residual specific heat. We hypothesize that the most likely interstitial is copper (a common impurity in tellurium). Our findings highlight the need to focus future order parameter determination efforts on single crystals of UTe$_{2}$ with minimal amounts of structural disorder.

\begin{acknowledgments}
    The authors acknowledge the significant contribution of the late Professor Lena F. Kourkoutis to this study.
    PFSR and CK acknowledge fruitful discussion with Eteri Svanidze and Juri Grin.
    Work at Los Alamos National Laboratory was performed under the auspices of the U.S. Department of
Energy, Office of Basic Energy Sciences, Division of Materials Science and Engineering. MMB acknowledges support from the Laboratory Directed Research and Development program, and CSK acknowledges support from the Seaborg Institute at Los Alamos.
This work made use of the electron microscopy facility of the Platform for the Accelerated Realization, Analysis, and Discovery of Interface Materials (PARADIM), which is supported by the National Science Foundation under Cooperative Agreement No. DMR-2039380, and the Cornell Center for Materials Research shared instrumentation facility.

\end{acknowledgments}

\subsection{Appendix}

\subsubsection{Experimental Methods}

UTe$_{2}$ single crystals were grown $via$ both chemical vapor transport and molten-salt-flux techniques, as described previously \cite{rosa2022,sakai2022}.

Specific heat was measured using the quasi-adiabatic thermal relaxation technique in a $^3$He cryostat insert. Both small-pulse and long-pulse methods were used to verify whether the samples had one or two transitions.

Scanning transmission electron microscopy (STEM) measurements were performed on cross-sectional lamellae prepared via gallium focused ion beam (FIB) on a Thermo Fisher Scientific Helios G4 UX FIB, with a 2~kV final thinning step. HAADF STEM imaging was performed on a Thermo Fisher Scientific Spectra 300 operating at 300~kV with a 30~mrad convergence semi-angle and 75~pA probe current. Series of rapid-frame images were acquired, aligned and averaged via a method of rigid registration optimized to prevent lattice hops to recover high signal-to-noise ratio, high fidelity atomic resolution images \cite{savitzky2018}. Interstitial sites with high HAADF-STEM signal, which arise from the presence of a high density of interstitials in the projection along a given axis, were first identified by successive filtering and thresholding and then further refined with manual inspection.

Single crystal X-ray diffraction experiments on samples s1 and s3 were performed using a Bruker D8 Venture. Single crystal X-ray diffraction experiments on sample s2 were performed using a Bruker D8 Quest. We note that there might be small systematic differences between the data collection of samples s1/s3 and the data collection of sample s2 because they were performed in different instruments. An Incoatec I$\mu$S microfocus source (Mo K-$\alpha$ radiation, $\lambda = 0.71073 \, \mathrm{\AA}$) was used as the radiation source. Data were collected using a PHOTON II CPAD area detector. 

The sample temperature was controlled by a constant flow of cold N$_2$ gas via Oxford Cryosystems N-Helix.
Data were collected between room temperature and 100~K. Each raw dataset was individually processed with Bruker SAINT software, including multi-scan absorption correction. The unit cell of each dataset was determined independently to quantify thermal contraction. The initial crystallographic model was obtained via the intrinsic phasing method in SHELXT. Least-squares refinements were performed using SHELXL2018 \cite{2015_Sheldrick_Acta}. 

\subsubsection{Oxide layer on STEM lamellas}

We note that the STEM lamellas for both samples s1 and s2 displayed a thick ($\sim 100-200$~nm) polycrystalline layer along their top edge, which was the only edge exposed to air prior to lifout. Electron energy loss spectrocopy shows that the polycrystalline layers contain uranium, tellurium, and oxygen. The interface between polycrystalline layers and the single crystal bulk of UTe$_{2}$ are not sharp, but the polycrystalline layer thickness is fairly consistent across the lamella. Thinner layers ($< 10$~nm) were also detected on the side facets ($i.e.,$ on the projection axis), which were only exposed to air after the lamellas were lifted out from the bulk. Because the lamellas were only exposed to air for a few hours between lifout and imaging, our results point to a fast degradation of UTe$_{2}$ surfaces.

\subsubsection{Data Availability}
The scanning transmission electron microscopy data as well as single crystal x-ray diffraction cif files are available at the Platform for the Accelerated Realization, Analysis, and Discovery of Interface Materials (PARADIM) database at https://doi.org/10.34863/xxxxx.


%

\end{document}